\documentclass[namedreferences]{kluwer}    

\usepackage{graphicx}

\newdisplay{guess}{Conjecture}

\begin{document}                                                                                   
\begin{article}
\begin{opening}         
\title{Advection dominated flows around black holes in two dimensions.} 
\author{William H. \surname{Lee}}  
\runningauthor{William Lee}
\runningtitle{ADAF}
\institute{Instituto de Astronom\'{\i}a, UNAM \\ Apdo. Postal 70--264,
Cd. Universitaria, M\'{e}xico D.F. 04510, M\'{e}xico}
\date{October 12, 2000}

\begin{abstract}
We show the results of two--dimensional, azimuthally symmetric
simulations of advection dominated fluid flows around black holes. We
use the SPH method, and the $\alpha$--prescription for the viscosity,
including all terms in the stress tensor. We have performed
calculations with $0.001 < \alpha < 0.2$ and find strong circulation
patterns in the flow, on spatial scales that depend on the magnitude
of the viscosity.
\end{abstract}

\end{opening}           

\section{Introduction, method and initial conditions}

The astrophysical relevance of accretion flows in which radiative
cooling is inefficient has been recently emphasized~\cite{NY94}. These
flows show complicated hydrodynamical behavior that is very different
from that obtained with the thin disk approach~\cite{SS73} for
efficient cooling.

We use the Smooth Particle Hydrodynamics (SPH) method~\cite{Monaghan},
in azimuthal symmetry. The equation of state is $P=\rho u (\gamma-1)$
with $\gamma=4/3$. Viscosity is computed using an $\alpha$
prescription ($\mu=\alpha \rho c_{s}^{2}/\Omega_{k}$), including all
the terms from the viscous stress tensor. The black hole is modeled as
a Newtonian point mass with an absorbing boundary at
$r_{in}=1.3r_{G}=1.3(2GM_{BH}/c^{2})$ . We use two kinds of initial
conditions: 1) equilibrium hydrostatic tori with constant specific
angular momentum; 2) matter injection at a steady rate at a radius
$r=300r_{G}$. We assume no radiative energy losses.

\section{Results}

\begin{figure}[H]
\tabcapfont
\centerline{%
\begin{tabular}{c@{\hspace{6pc}}c}
\includegraphics[width=1.7in,angle=-90]{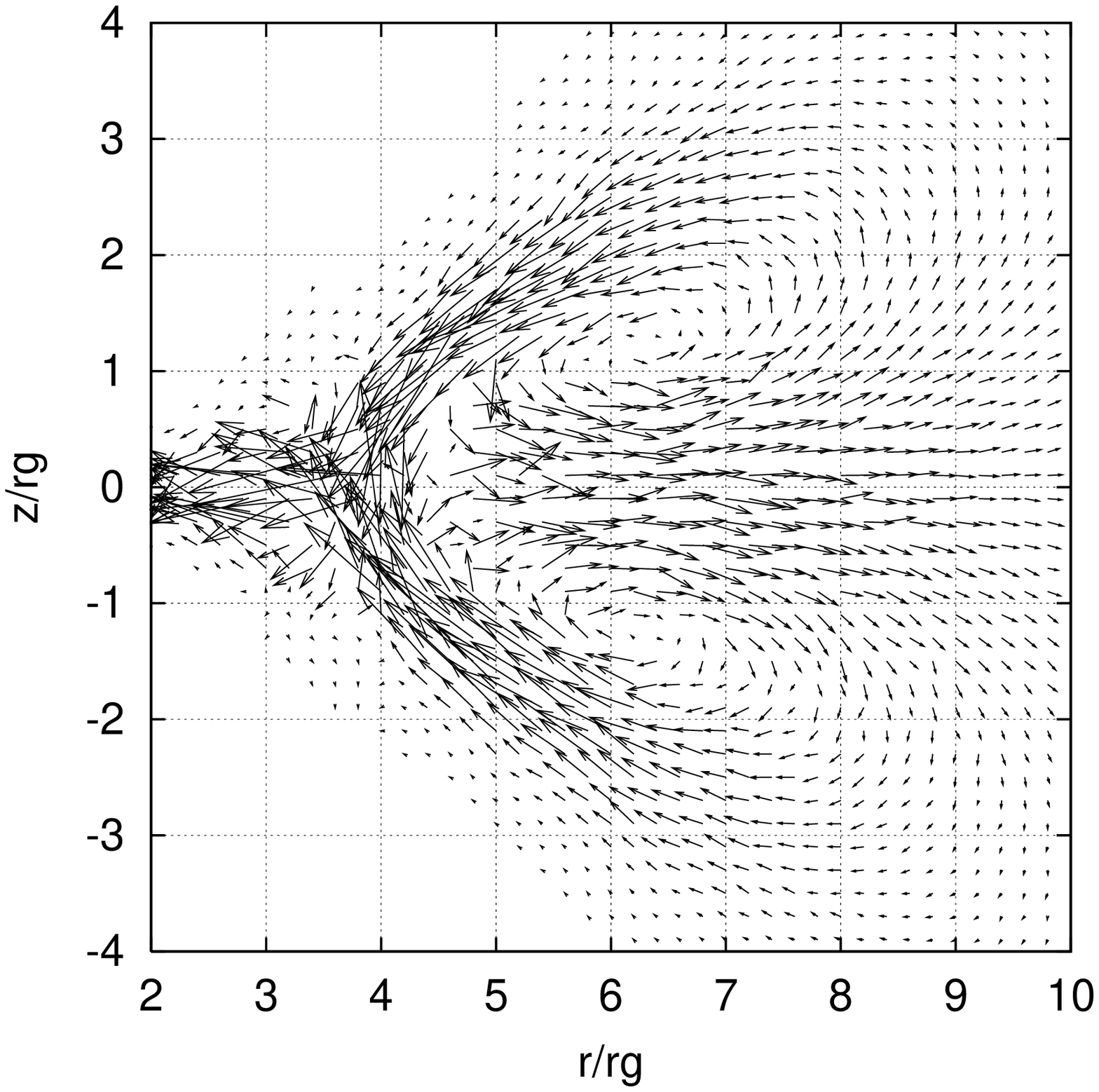} &
\includegraphics[width=1.7in,angle=-90]{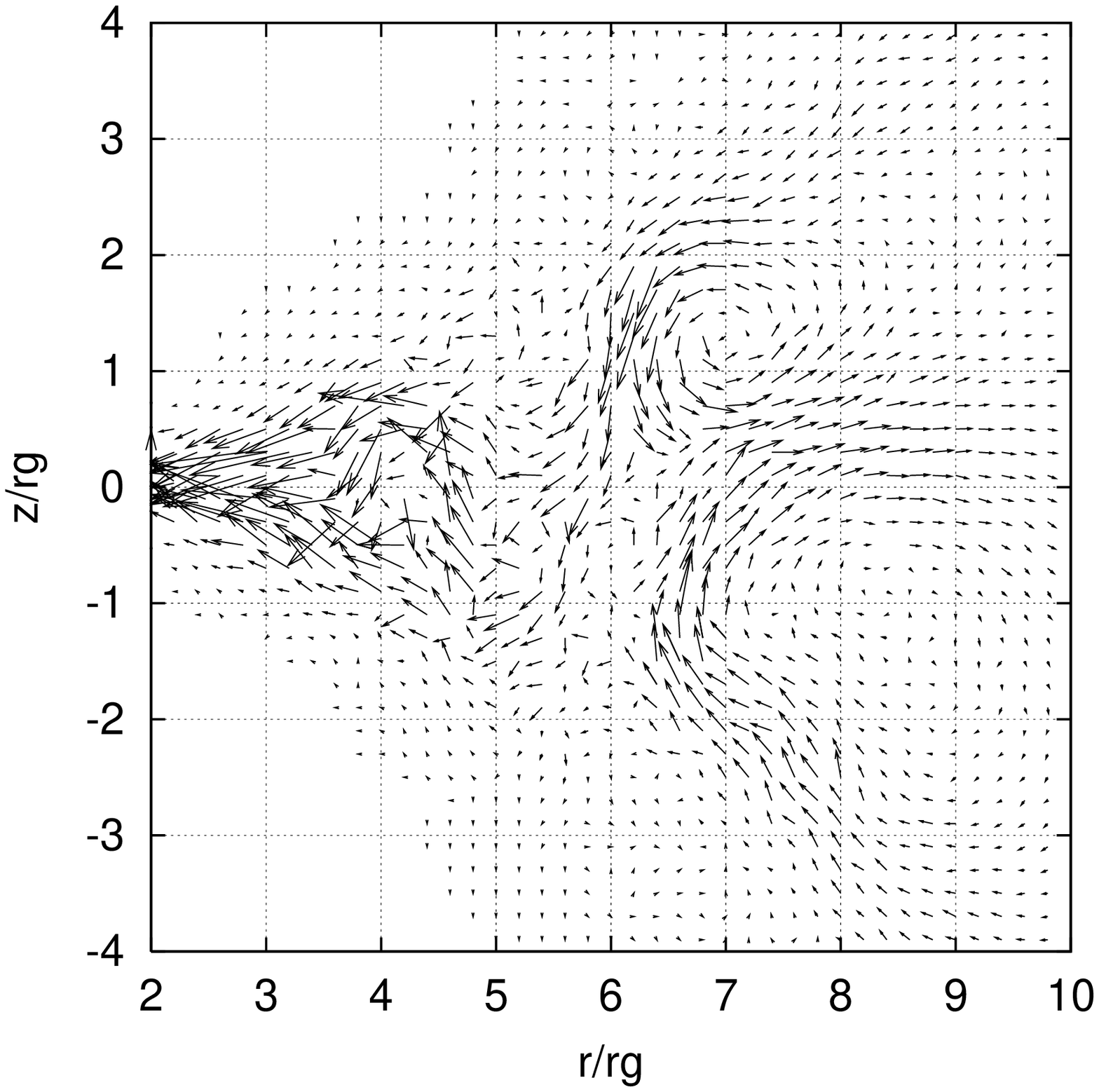} \\
a.~~ $t=200r_{G}/c$ & b.~~ $t=800r_{G}/c$
\end{tabular}}
\caption{Momentum vector field $\rho \vec{v}$ for $\alpha=0.01$. }\label{mom}
\end{figure}

For initially hydrostatic tori, the presence of a hot, high--entropy
bubble near the equatorial plane is easily seen
(Figure~\ref{mom}). This moves outward in the disk, on top of a
background accretion flow. There is a large--scale circulation
pattern, on a scale $z\simeq r$. For a lower viscosity $\alpha=0.001$
there are small--scale vortices in the disk, and the overall behavior
is more turbulent. A background accretion flow is also present.

For the case of matter injection, the flow structure is governed by
the magnitude of the viscosity, with small--scale eddies at low values
of $\alpha$ and large--scale circulations when the viscosity is
high. The fluid shows significant oscillations in the vertical
direction as it moves radially inward.

With these simulations we have tested our code with previously
published results obtained with a different numerical
scheme~\cite{igu96,igu99}, and we have found excellent agreement both
in a qualitative and quantitative manner. In future work we will
extend our results to study the parameter space ($\alpha$, $\gamma$)
and increase the resolution.

\acknowledgements I acknowledge support from CONACyT (27987E),
DGAPA (IN-119998) and helpful conversations with M. Abramowicz
and W. Klu\'{z}niak.

\end{article}
\end{document}